\documentclass{aip-cp}

\usepackage[numbers]{natbib}
\usepackage{rotating}
\usepackage{graphicx}
\usepackage{multirow}

\begin{document}

\title{Studies of Charmonium at BESIII}

\author[aff1]{X.~C.~Ai\corref{cor1} on behalf of the BESIII Collaboration}
\eaddress[url]{http://www.aip.org}

\affil[aff1]{Institute of High Energy Physics, Chinese Academy of Sciences}
\corresp[cor1]{Corresponding author: aixc@ihep.ac.cn}

\maketitle

\begin{abstract}
In recent years, lots of studies of charmonium decays have been performed at BESIII based on large data samples of $J/\psi$, $\psi(3686)$ and $\psi(3770)$. Recent results in searches for radiative transitions of $\psi(3770)$ and rare phenomena in charmonium decays, and studies of light hadrons structures and properties will be presented.
\end{abstract}

\section{INTRODUCTION}
The BESIII detector at the BEPCII collider has accumulated the world's largest data samples of $e^+e^-$ collisions in the $\tau$-charm region, including $J/\psi$, $\psi(3686)$, $\psi(3770)$ and the XYZ data. These data samples provide a clean and simple environment to study the production and decay mechanisms of charmonium states including not only $J/\psi$, $\psi(3686)$, $\psi(3770)$, but also $\eta_c$, $\eta_c(2S)$ and $\chi_{cJ}$ which are available by the $\gamma$ transition of $\psi(3686)$ and $\psi(3770)$.

Results in this presentation are based on 2.92 fb$^{-1}$ data sample taken at $\sqrt{s}$ = 3.773 GeV, 1.06 $\times$ $10^8$ $\psi(3686)$ events and 1.31 $\times$ $10^9$ $J/\psi$ events.

\section{SEARCH FOR $\psi(3770)$ RADIATIVE TRANSITIONS}
The nature of the excited $J^{PC}$ = $1^{--}$ $c\bar{c}$ bound states above the $D\bar{D}$ threshold is of interest but still not well known. The $\psi(3770)$ resonance, as the lightest charmonium state lying above the open charm threshold, is
generally assigned to be a dominant $1^3 D_1$ momentum
eigenstate with a small $2^3 S_1$ admixture~\cite{bib1}. It has been
thought almost entirely to decay to $D\bar{D}$ states~\cite{bib2,bib3}. Unexpectedly, the BES Collaboration found a large inclusive non-$D\bar{D}$ branching fraction, (14.7 $\pm$ 3.2)\%, by utilizing various methods~\cite{bib4,bib5,bib6,bib7}. A
later work by the CLEO Collaboration found a contradictory non-$D\bar{D}$ branching fraction,(3.3 $\pm$ 1.4$_{-4.8}^{+6.6}$)\%~\cite{bib8}. The BES results suggest substantial non-$D\bar{D}$ decays, although the CLEO
result finds otherwise. In the exclusive analyses, various non-$D\bar{D}$ decay
modes have been observed, including hadronic transitions $\psi(3770)\rightarrow J/\psi\pi^+\pi^-$~\cite{bib9,bib10}, $\pi^0\pi^0J/\psi$, $\eta J/\psi$ ~\cite{bib10}, the $E1$ transitions $\gamma \chi_{cJ}$ (J = 0, 1)~\cite{bib11,bib12}, and the decay to light hadrons $\phi\eta$~\cite{bib13}. The sum of the observed non-$D\bar{D}$ exclusive components still makes up less than 2\%
of all decays~\cite{pdg}, which motivates the search for other
exclusive non-$D\bar{D}$ final states.

\subsection{Search for $\psi(3770)\rightarrow\gamma\eta_c$ and $\gamma\eta_c(2S)$}
The radiative transitions $\psi(3770)\rightarrow\gamma\eta_c(\eta_c(2S))$ are supposed to be highly suppressed by selection rules, considering
the $\psi(3770)$ is predominantly the $1^3D_1$ state. However, due
to the non-vanishing photon energy in the decay, higher
multipoles beyond the leading one could contribute~\cite{bib15}. Recently, the partial
decay widths $\Gamma(\psi(3770)\rightarrow\gamma\eta_c(\eta_c(2S))$ have been calculated in Ref.~\cite{bib15} by considering contributions from the
intermediate meson loop (IML) mechanism. 

\begin{figure}[h]
  \includegraphics[width=200pt]{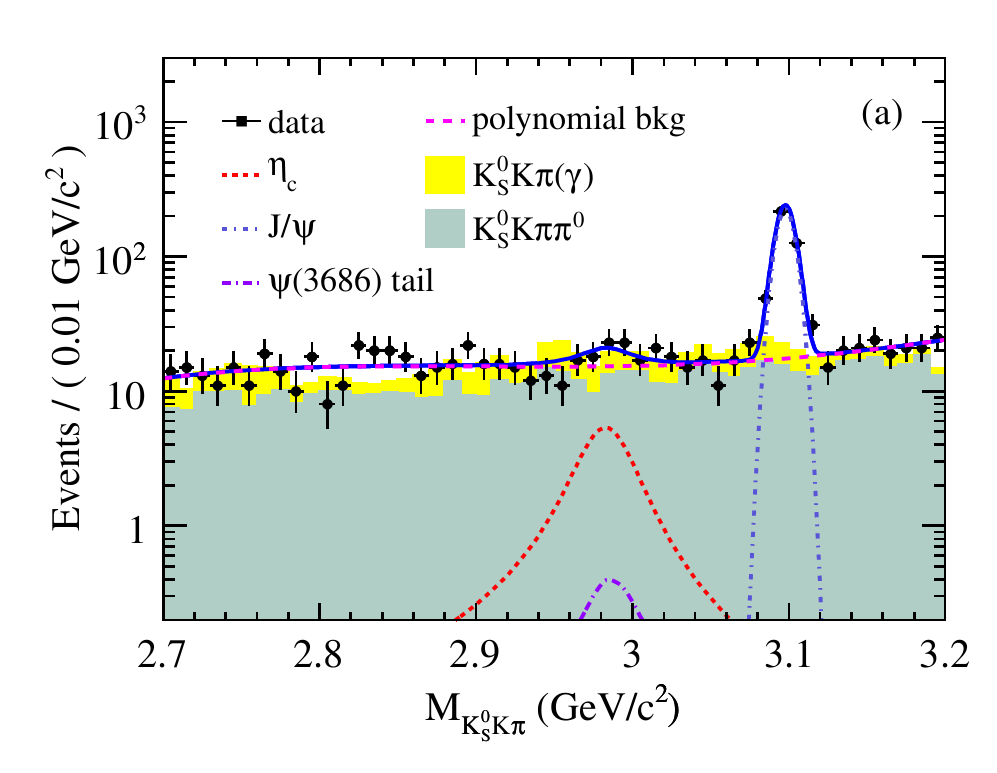}
  \includegraphics[width=200pt]{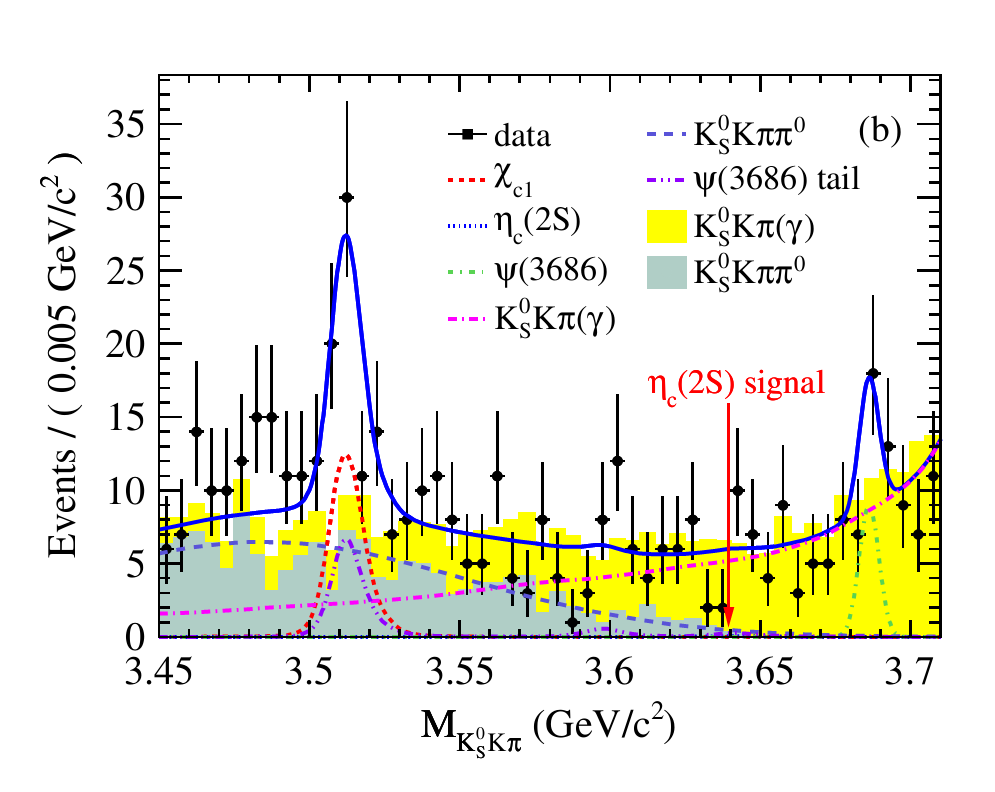}
  \caption{Invariant-mass spectrum for $K_S^0 K^{\pm}\pi^{\mp}$ from data with the estimated backgrounds and best-fit results
superimposed in the (a)~$\eta_c$ and (b)~$\chi_{c1}$ and $\eta_c(2S)$ mass regions. Dots with error bars are data. The shaded histograms represent the
background contributions and the solid lines show the total fit results.}
   \label{fig:mkskpi}
\end{figure}

Using the $\psi(3770)$ data sample, the radiative transitions $\psi(3770)\rightarrow\gamma\eta_c(\eta_c(2S))$ through the decay process $\psi(3770)\rightarrow K_S^0 K^{\pm}\pi^{\mp}$ have been searched for~\cite{bibetac}. Figure ~\ref{fig:mkskpi} shows the invariant-mass spectrum of $K_S^0 K^{\pm}\pi^{\mp}$ for selected candidates, together with the estimated backgrounds in the $\eta_c$
mass region (Fig.~\ref{fig:mkskpi}(a)) and in the $\chi_{c1}$ $-$ $\eta_c(2S)$ mass region (Fig.~\ref{fig:mkskpi}(b)). No significant $\eta_c$ and $\eta_c(2S)$ signals are observed. The upper limits on
the branching fractions at a 90\% C.L. have been set: ${\mathcal B}(\psi(3770)\rightarrow\gamma\eta_c) < 6.8 \times 10^{-4}$ and ${\cal B}(\psi(3770)\rightarrow\gamma\eta_c(2S)) < 2.0 \times 10^{-3}$. The upper limit for $\Gamma(\psi(3770)\rightarrow\gamma\eta_c)$ is within the error
range of the theoretical predictions of IML~\cite{bib15} and lattice QCD calculations~\cite{bib16}. However, the upper
limit for $\Gamma(\psi(3770)\rightarrow\gamma\eta_c(2S))$ is much larger than the
prediction and is limited by statistics and the systematic error.

\subsection{Measurement of the Branching Fraction for $\psi(3770)\rightarrow\gamma\chi_{cJ}$}
Within an S-D mixing model~\cite{bib1}, Refs.~\cite{bib17,bib18,bib19} predict the partial widths for $\psi(3770)$ $E1$
radiative transitions, but with large uncertainties. Up to now, the transition $\psi(3770)\rightarrow\gamma\chi_{c2}$ has not been observed. Precision measurements of partial widths of the $\psi(3770)\rightarrow\gamma\chi_{c1,2}$ processes are critical
to test the theoretical predictions, and to better understand
the nature of the partial widths of the $\psi(3770)$, as well as to find the origin of the
non-$D\bar{D}$ decays of $\psi(3770)$.

\begin{figure}[h]
  \includegraphics[width=200pt]{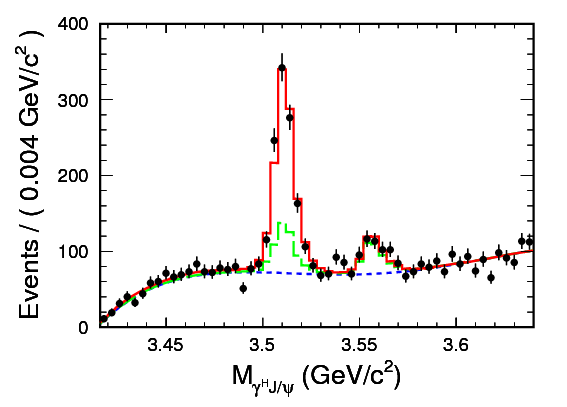}
  \caption{Invariant mass spectrum of higher energetic photon and $J/\psi$ selected from data. The dots with error bars
represent the data. The solid (red) line shows the fit. The dashed
(blue) line shows the smooth background. The long-dashed
(green) line is the sum of the smooth background and the
contribution from $e^+e^-\rightarrow(\gamma_{ISR})\psi(3686)$ production.}
   \label{fig:mgjpsi}
\end{figure}

Using the $\psi(3770)$ data sample, the $\psi(3770)$ $E1$ transitions $\psi(3770)\rightarrow\gamma\chi_{c1,2}$ have been studied~\cite{bibchicJ} by reconstructing $\chi_{cJ}$ using the decay $\chi_{cJ}\rightarrow\gamma J/\psi$. Figure~\ref{fig:mgjpsi} shows the invariant-mass distribution of the higher energetic photon and $J/\psi$. Clear peak corresponding to
the $\chi_{c1}$ signal is observed while there is no significant signal of $\chi_{c2}$. The branching fraction of $\psi(3770)\rightarrow\gamma\chi_{c1}$ is measured to be ${\cal B}(\psi(3770)\rightarrow\gamma\chi_{c1}) = (2.48\pm0.15\pm0.23)\times 10^{-3}$ , which is the most precise measurement to date. The upper limit on the branching fraction of $\psi(3770)\rightarrow \gamma\chi_{c2}$ at 90\% C.L. is ${\cal B}(\psi(3770)\rightarrow\gamma\chi_{c2})<0.64\times10^{-3}$.

\section{SEARCH FOR RARE PHENOMENA IN CHARMONIUM DECAYS}
\subsection{Search for Isospin-violating Transition $\chi_{c0,2}\rightarrow\pi^0\eta_c$}
Based upon an effective-field theoretical approach, theoretical calculations give qualitative insights in 
the isospin-breaking mechanisms in charmonium decays below $D\bar{D}$ threshold~\cite{bib25}. Currently, for such a theory, 
quantitative predictions of individual
branching fractions of isospin-forbidden decays of charmonium require more constraints
from experimental data.

Using the $\psi(3686)$ data sample, the hadronic isospin-violating transitions $\chi_{c0,2}\rightarrow\pi^0\eta_c$ have been searched for through 
$\eta_c\rightarrow K_S^0 K^{\pm}\pi^{\mp}$ decays~\cite{bibpi0etac}. No statistically significant
signal is observed and upper limits on the branching
fractions for the processes $\chi_{c0,2}\rightarrow\pi^0\eta_c$ have been
obtained. The results are ${\cal B}(\chi_{c0}\rightarrow \pi^0\eta_c) < 1.6\times 10^{-3}$ and ${\cal B}(\chi_{c2}\rightarrow \pi^0\eta_c) < 3.2\times 10^{-3}$.
These are the first upper limits on ${\cal B}(\chi_{c0,2}\rightarrow \pi^0\eta_c)$ that have been reported so far. These limits might
help to constrain non-relativistic field theories and provide
insight in the role of charmed-meson loops to the various
transitions in charmonium and charmonium-like states.
 
\subsection{Search for $C$-parity Violation in $J/\psi\rightarrow\gamma\gamma$, $\gamma\phi$} 
In the Standard Model (SM), $C$ invariance is held in strong and electromagnetic (EM) interactions. Until now, no $C$-violating processes have been observed in EM interactions. Any evidence for $C$ violation in the EM sector would immediately indicate physics beyond the SM.

Using the $\psi(3686)$ data sample, the decays of $J/\psi\rightarrow\gamma\gamma$ and $\gamma\phi$ have been searched for via $\psi(3686)\rightarrow J/\psi\pi^+\pi^-$~\cite{bibgggphi}. No significant signal is observed for $J/\psi\rightarrow\gamma\gamma$ and $J/\psi\rightarrow\gamma\phi$. The upper limits for the branching fractions of $J/\psi\rightarrow \gamma\gamma$ and $J/\psi\rightarrow\gamma\phi$ are set to be ${\cal B}(J/\psi\rightarrow \gamma\gamma) < 2.7 \times 10^{-7}$ and ${\cal B}(J/\psi\rightarrow \gamma\phi) < 1.4 \times 10^{-6}$, respectively.
The upper limit on ${\cal B}(J/\psi\rightarrow \gamma\gamma)$ is one of magnitude more stringent
than the previous upper limit~\cite{bib26}, and ${\cal B}(J/\psi\rightarrow \gamma\phi)$ is the first upper limit for this channel.

\subsection{Observation of OZI-suppressed Decay $J/\psi \rightarrow \pi^0 \phi$}
A full investigation of $J/\psi$ decaying to a vector meson
($V$) and a pseudoscalar meson ($P$) can provide rich
information about SU(3) flavor symmetry and its breaking,
probe the quark and gluon content of the pseudoscalar
mesons, and determine the electromagnetic amplitudes~\cite{bib27,bib28,bib29}. 

\begin{figure}[h]
    \includegraphics[width=200pt]{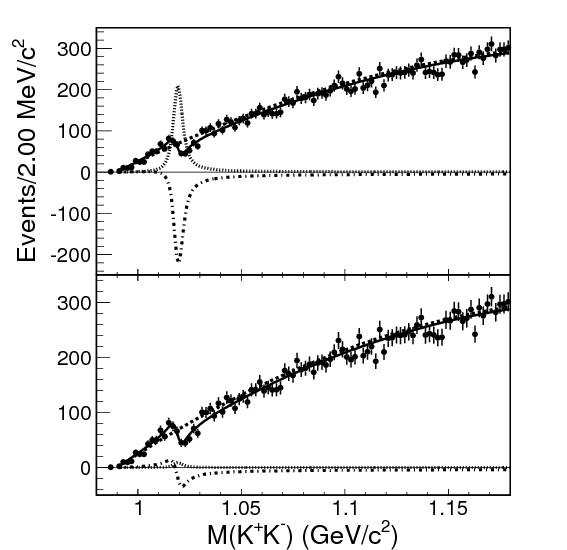}                                                                                                                                                                                           \caption{Fit to $M(K^+K^-)$ spectrum after sideband subtraction for Solution I (a) and Solution II (b). The red dotted
curve denotes the $\phi$ resonance; the blue dashed curve is the non-$\phi$ contribution; the green dot-dashed curve represents their inter-ference; the blue solid curve is the sum of them.}
     \label{fitmkk}                                                                                                                              
   \end{figure}

Using the $J/\psi$ data sample, the first evidence for a doubly OZI suppressed electromagnetic $J/\psi$ decay $J/\psi\rightarrow \pi^0\phi \rightarrow K^+K^-\gamma\gamma$ has been reported~\cite{bib30}. A clear structure in the $K^+K^-$ invariant mass spectrum around 1.02 GeV/$c^2$ is observed, which can be attributed to interference of $J/\psi\rightarrow \pi^0\phi$ and $J/\psi\rightarrow K^+K^-\pi^0$ decays. Figure~\ref{fitmkk} shows the fit to the invariant mass spectrum of $K^+K^-$ with the background events estimated with $\pi^0$ sidebands subtracted. Due to the interference, two possible solutions are found. The corresponding measured
values of the branching fraction of $J/\psi \rightarrow \pi^0 \phi$ are ($2.94\pm0.16\pm0.16$) $\times$ $10^{-6}$ and ($1.24\pm0.33\pm0.30$) $\times$ $10^{-7}$, respectively.
  
\section{STUDIES OF LIGHT HADRON STRUCTURES AND PROPERTIES}
\subsection{Study of $J/\psi\rightarrow \phi \pi^0 f_0(980)$}
The nature of the scalar meson $f_0(980)$ is a long-standing puzzle~\cite{pdg}. In the study of $J/\psi$ radiatively decaying into $\pi^+\pi^-\pi^0$ and $\pi^0\pi^0\pi^0$, BESIII observed the decay of $\eta(1405)\rightarrow\pi^0 f_0(980)$ with a large isospin violation and an anomalously narrow width of $f_0(980)$~\cite{bib1405}. One proposed explanation for these phenomena is the triangle singularity mechanism~\cite{tsm1, tsm2}.

\begin{figure}[!htp]
  \includegraphics[width=200pt]{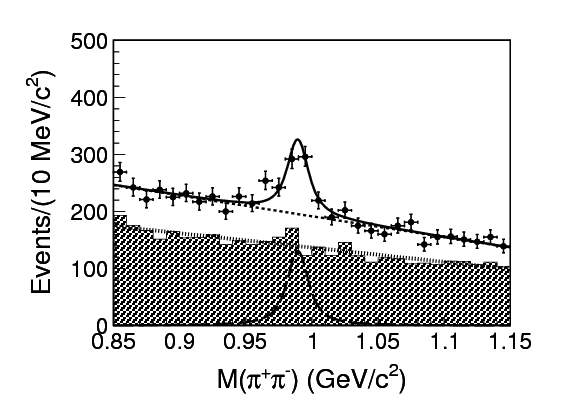}
  \put(-35,115){(a)}
  \includegraphics[width=200pt]{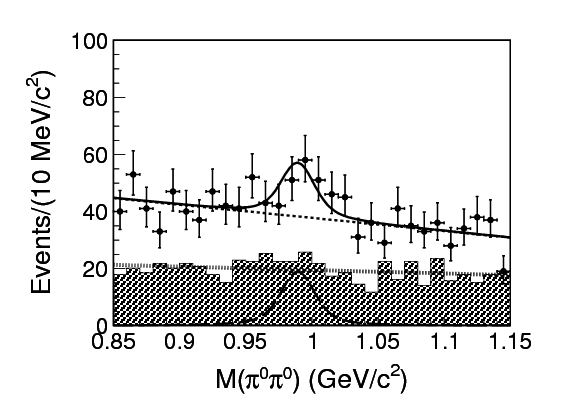}
  \put(-35,115){(b)}
  \caption{The spectra (a)~$M(\pi^+\pi^-)$ and (b)~$M(\pi^0\pi^0)$ (three entries
per event) with $K^+K^-$ in the $\phi$ signal region (black dots) and in the $\phi$ sideband regions (hatched histogram). The solid curve is the full fit, the long-dashed curve is the $f_0(980)$ signal, the dotted line is the
non-$\phi$ background, and the short-dashed line is the total background.}
   \label{fig:mpipi_980}
 \end{figure}

Using the $J/\psi$ data sample, the decays $J/\psi\rightarrow \phi\pi^0\pi^0\pi^0$ with $\phi\rightarrow K^+K^-$ are investigated~\cite{bib14052}. The isospin-violating decay $J/\psi\rightarrow \phi \pi^0 f_0(980)$ is observed for the first time. Figure~\ref{fig:mpipi_980} shows the invariant mass spectrum of $\pi^+\pi^-$ (Fig.~\ref{fig:mpipi_980}(a)) and $\pi^0\pi^0$ (Fig.~\ref{fig:mpipi_980}(b)). A clear $f_0(980)$ exists for the $\pi^+\pi^-$ mode. The width obtained from the dipion mass spectrum is ($15.3\pm4.7$) MeV/$c^2$, which is consistent with that in the study of $J/\psi\rightarrow \gamma \eta(1405)\rightarrow \gamma \pi^0 f_0(980)$~\cite{bib1405} and the prediction of a theoretical work~\cite{bibtsm3} based on the triangle singularity mechanism~\cite{tsm1, tsm2}. In the invariant mass spectra of $f_0(980)\pi^0$, there is evidence of a resonance around 1.28 GeV/$c^2$ for the $f_0(980)\rightarrow \pi^+\pi^-$ mode, which was identified as the axial-vector meson $f_1(1285)$.

\subsection{Study of $\chi_{cJ}\rightarrow \eta'K^+K^-$}
Until now, $K_0^*(1430)$ has
been observed in $K_0^*(1430)\rightarrow K\pi$ only, but it is also
expected to couple to $\eta'K$~\cite{etapk1, etapk2}. $\chi_{c1}\rightarrow\eta'K^+K^-$ is a promising channel to search for $K_0^*(1430)$ and study its properties while decays of $\chi_{c0,2}\rightarrow K_0^*(1430)K$ are forbidden
by spin-parity conservation.

\begin{figure*}[h]
 \centering
  \includegraphics[width=200pt]{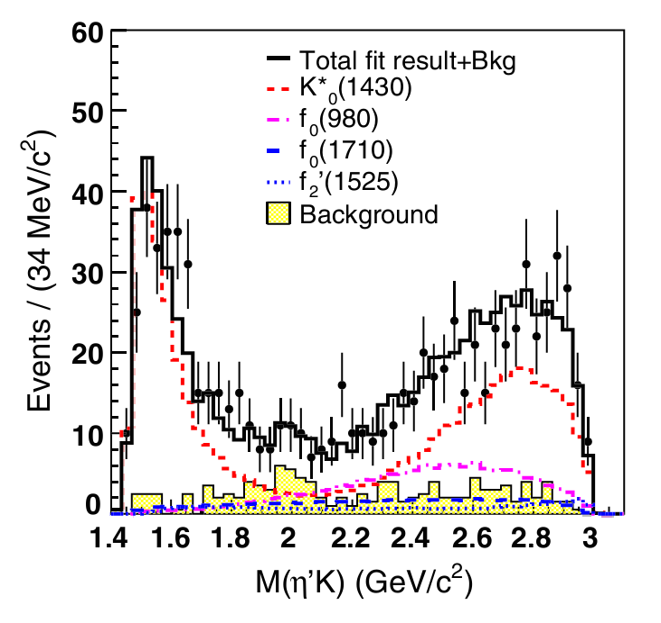}
  \put(-30,155){(a)}
  \includegraphics[width=200pt]{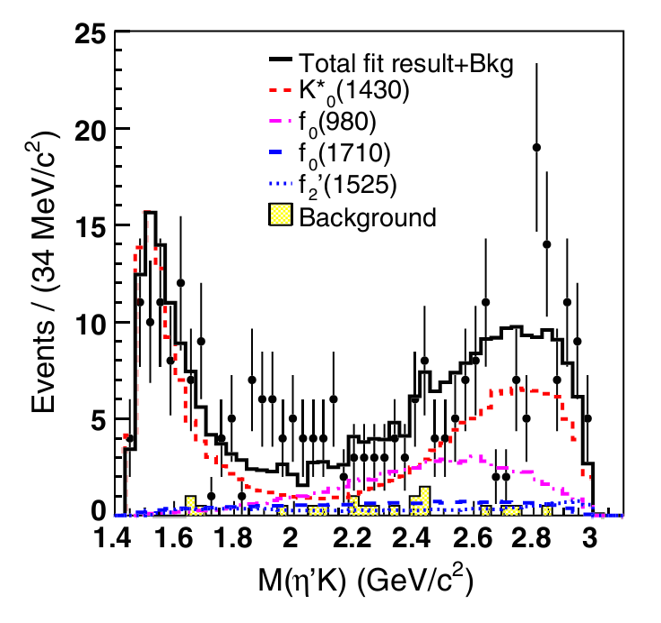}
  \put(-30,155){(b)}
  \caption{The invariant mass distribution of $\eta'K^{\pm}$ within the $\chi_{c1}$ mass range for (a)~$\eta'\rightarrow\gamma \rho^0$ mode and (b)~$\eta'\rightarrow\eta\pi^{+}\pi^{-}$ mode.}
   \label{fig:metapk}
 \end{figure*}

Using the $\psi(3686)$ data sample, the decay $\chi_{cJ}\rightarrow \eta'K^+K^-$ with $\eta'\rightarrow \gamma \rho^0$ and $\eta'\rightarrow\eta\pi^{+}\pi^{-}$, $\eta\rightarrow\gamma\gamma$ is studied for the first time~\cite{etapkk}. Abundant structures on the $K^+K^-$ and $\eta K^{\pm}$ invariant mass spectra are observed for $\chi_{c1}$ candidate events, and a partial wave analysis is performed for the decay $\chi_{c1}\rightarrow \eta'K^+K^-$. Figure~\ref{fig:metapk} shows the comparisons of data and fit projections in terms of the invariant mass spectra of $\eta'K^{\pm}$ for $\chi_{c1}$ candidate events. The partial branching fractions of $\chi_{c1}$ decay processes with intermediate states $f_0(980)$, $f_0(1710)$, $f_2'(1525)$ and $K_0^*(1430)$ are measured for the first time.
%

\subsection{Study of $\chi_{cJ}$ Decaying into $\phi K^*(892)\bar{K}$}
The nature of the axial-vector candidate, $h_1(1380)$ is still controversial~\cite{13801,13802}. The direct observation of the $h_1(1380)$ in experiments and the precise measurement of its resonance parameters may shed light on its nature and aid in identifying the ground state axial-vector meson nonet in the
quark model.

\begin{figure}[!htp]
  \includegraphics[width=200pt]{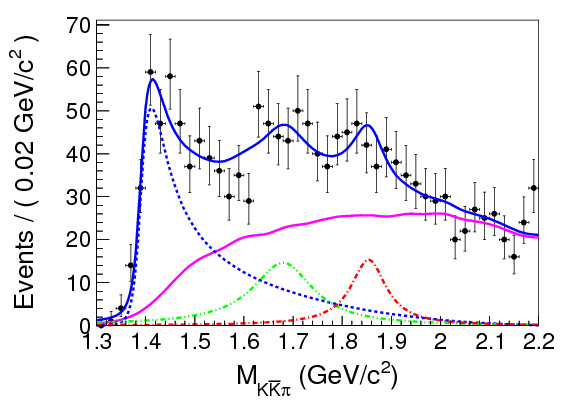}
  \caption{The sum of $K_S^0 K^{\pm} \pi^{\mp}$ and $K^{\pm} \pi^{\mp}\pi^0$ mass
spectra in the $\chi_{c1}$ and $\chi_{c2}$ mass regions. The markers with error
bars represent the data; the dash curve the $h_1(1380)$ signal; the
dash-dot-dot curve the $\phi(1680)$ signal; and the dash-dot curve the $\phi(1850)$ signal.}
   \label{fig:1380}
 \end{figure}

Using the $\psi(3686)$ data sample, the first measurement of $\chi_{cJ}\rightarrow \phi K_S^0 K^{\pm} \pi^{\mp}$ and 
$\chi_{cJ}\rightarrow \phi K^+ K^- \pi^0$ has been reported~\cite{k892}. The decays are dominated by the three-body reaction $\chi_{cJ}$ decaying into $\phi K^*(892)\bar{K}$. The branching
fractions for this reaction via neutral and charged $K^*(892)$ are measured for the first time. Figure~\ref{fig:1380} shows the invariant mass spectrum of $K\bar{K}\pi$, a
significant excess of events above the phase space expectation is observed near the $K^*(892)\bar{K}$ mass threshold in the
decays of $\chi_{c1,2}$ with a significance greater than $10\sigma$. The
observed structure has negative $C$ parity, and is expected to
be the $h_1(1380)$ state, considering its mass, width and decay through $K^*(892)\bar{K}$. The mass and width
of the $h_1(1380)$ are determined to ($1412\pm4\pm8$) MeV/c$^2$ and ($84\pm12\pm40$) MeV, respectively. This is the first evidence of the $h_1(1380)$ in its decay to $K^*(892)\bar{K}$. Evidence is also found
for the decays $\chi_{cJ}\rightarrow \phi\phi(1680)$ and $\chi_{cJ}\rightarrow \phi\phi(1850)$,but
with significances less than $5\sigma$.
%

\section{SUMMARY}
Based on 2.92 fb$^{-1}$ data sample taken at $\sqrt{s}$ = 3.773 GeV, 1.06$\times10^8$ $\psi(3686)$ events and 1.31$\times10^9$ $J/\psi$ events collected with BESIII detector at the BEPCII collider, studies have been performed to search for the radiative transitions of $\psi(3770)$ and rare decays of charmonium, and study light hadron structures and properties. The upper limits on branching fractions are set for radiative transitions of $\psi(3770)$, $\psi(3770)\rightarrow \gamma \eta_c(\eta_c(2S))$ and $\psi(3770)\rightarrow \gamma \chi_{c2}$, the isospin-violating decay $\chi_{c0,2}\rightarrow\pi^0\eta_c$, and $C$-violation decay $J/\psi\rightarrow\gamma\gamma$, $\gamma\phi$. The OZI-suppressed decay $J/\psi\rightarrow \pi^0\phi$ is observed for the first time. The isospin-violating decay $J/\psi\rightarrow \phi\pi^0 f_0(980)$, $f_0(980)\rightarrow\pi^+\pi^-$ is observed with the width of the $f_0(980)$ obtained
from the dipion mass spectrum found to be much smaller than the world average value. The decay $\chi_{c1,2}\rightarrow \eta'K^+K^-$ is studied for the first time. Intermediate process $\chi_{c1}\rightarrow \eta'f_0(980)$, $\chi_{c1}\rightarrow\eta'f_0(1710)$, $\chi_{c2}\rightarrow\eta'f_2'(1525)$ and $\chi_{c1}\rightarrow
K_0^*(1430)^{\pm}K^{\mp}$ are observed and their branching fractions are measured. The branching fractions of decay $\chi_{cJ}$ decaying into $\phi K^*(892)\bar{K}$ are measured for the first time and the first evidence of the $h_1(1380)$ in its decay to $K^*(892)\bar{K}$ is obtained. These measurements provide important information to understand the nature of $\psi(3770)$, the isospin-violation mechanism and properties of light hadrons, such as $f_0(980)$, $K_0^*(1430)$ and $h_1(1380)$.


\nocite{*}
\bibliographystyle{aipnum-cp}%
\bibliography{proceeding}%

\begin{thebibliography}{40}%
\makeatletter
\providecommand \@ifxundefined [1]{%
 \@ifx{#1\undefined}
}%
\providecommand \@ifnum [1]{%
 \ifnum #1\expandafter \@firstoftwo
 \else \expandafter \@secondoftwo
 \fi
}%
\providecommand \@ifx [1]{%
 \ifx #1\expandafter \@firstoftwo
 \else \expandafter \@secondoftwo
 \fi
}%
\providecommand \natexlab [1]{#1}%
\providecommand \enquote  [1]{``#1''}%
\providecommand \bibnamefont  [1]{#1}%
\providecommand \bibfnamefont [1]{#1}%
\providecommand \citenamefont [1]{#1}%
\providecommand \href@noop [0]{\@secondoftwo}%
\providecommand \href [0]{\begingroup \@sanitize@url \@href}%
\providecommand \@href[1]{\@@startlink{#1}\@@href}%
\providecommand \@@href[1]{\endgroup#1\@@endlink}%
\providecommand \@sanitize@url [0]{\catcode `\$12\catcode `\&12\catcode
  `\#12\catcode `\^12\catcode `\_12\catcode `\%12\relax}%
\providecommand \@@startlink[1]{}%
\providecommand \@@endlink[0]{}%
\providecommand \url  [0]{\begingroup\@sanitize@url \@url }%
\providecommand \@url [1]{\endgroup\@href {#1}{\urlprefix }}%
\providecommand \urlprefix  [0]{URL }%
\providecommand \Eprint [0]{\href }%
\providecommand \doibase [0]{http://dx.doi.org/}%
\providecommand \selectlanguage [0]{\@gobble}%
\providecommand \bibinfo  [0]{\@secondoftwo}%
\providecommand \bibfield  [0]{\@secondoftwo}%
\providecommand \translation [1]{[#1]}%
\providecommand \BibitemOpen [0]{}%
\providecommand \bibitemStop [0]{}%
\providecommand \bibitemNoStop [0]{.\EOS\space}%
\providecommand \EOS [0]{\spacefactor3000\relax}%
\providecommand \BibitemShut  [1]{\csname bibitem#1\endcsname}%
\let\auto@bib@innerbib\@empty
\bibitem [{\citenamefont {Rosner}(2001{\natexlab{a}})}]{bib1}%
  \BibitemOpen
  \bibfield  {author} {\bibinfo {author} {\bibfnamefont {J.~L.}\ \bibnamefont
  {Rosner}},\ }\href@noop {} {\bibfield  {journal} {\bibinfo  {journal} {Phys.
  Rev. D}\ }\textbf {\bibinfo {volume} {64}},\ p.\ \bibinfo {pages} {094002}
  (\bibinfo {year} {2001}{\natexlab{a}})}\BibitemShut {NoStop}%
\bibitem [{\citenamefont {Rapidis}\ \emph {et~al.}(1977)\citenamefont {Rapidis}
  \emph {et~al.}}]{bib2}%
  \BibitemOpen
  \bibfield  {author} {\bibinfo {author} {\bibfnamefont {P.~A.}\ \bibnamefont
  {Rapidis}} \emph {et~al.},\ }\href@noop {} {\bibfield  {journal} {\bibinfo
  {journal} {Phys. Rev. Lett.}\ }\textbf {\bibinfo {volume} {39}},\ p.\
  \bibinfo {pages} {526} (\bibinfo {year} {1977})}\BibitemShut {NoStop}%
\bibitem [{\citenamefont {Bacino}\ \emph {et~al.}(1978)\citenamefont {Bacino}
  \emph {et~al.}}]{bib3}%
  \BibitemOpen
  \bibfield  {author} {\bibinfo {author} {\bibfnamefont {W.}~\bibnamefont
  {Bacino}} \emph {et~al.},\ }\href@noop {} {\bibfield  {journal} {\bibinfo
  {journal} {Phys. Rev. Lett.}\ }\textbf {\bibinfo {volume} {40}},\ p.\
  \bibinfo {pages} {671} (\bibinfo {year} {1978})}\BibitemShut {NoStop}%
\bibitem [{\citenamefont {Ablikim}\ \emph {et~al.}(2007)\citenamefont {Ablikim}
  \emph {et~al.}}]{bib4}%
  \BibitemOpen
  \bibfield  {author} {\bibinfo {author} {\bibfnamefont {M.}~\bibnamefont
  {Ablikim}} \emph {et~al.} (\bibinfo {collaboration} {BES Collaboration}),\
  }\href@noop {} {\bibfield  {journal} {\bibinfo  {journal} {Phys. Rev. D}\
  }\textbf {\bibinfo {volume} {76}},\ p.\ \bibinfo {pages} {122002} (\bibinfo
  {year} {2007})}\BibitemShut {NoStop}%
\bibitem [{\citenamefont {Ablikim}\ \emph {et~al.}(2008)\citenamefont {Ablikim}
  \emph {et~al.}}]{bib5}%
  \BibitemOpen
  \bibfield  {author} {\bibinfo {author} {\bibfnamefont {M.}~\bibnamefont
  {Ablikim}} \emph {et~al.} (\bibinfo {collaboration} {BES Collaboration}),\
  }\href@noop {} {\bibfield  {journal} {\bibinfo  {journal} {Phys. Lett. B}\
  }\textbf {\bibinfo {volume} {659}},\ p.~\bibinfo {pages} {74} (\bibinfo
  {year} {2008})}\BibitemShut {NoStop}%
\bibitem [{\citenamefont {Ablikim}\ \emph
  {et~al.}(2006{\natexlab{a}})\citenamefont {Ablikim} \emph {et~al.}}]{bib6}%
  \BibitemOpen
  \bibfield  {author} {\bibinfo {author} {\bibfnamefont {M.}~\bibnamefont
  {Ablikim}} \emph {et~al.} (\bibinfo {collaboration} {BES Collaboration}),\
  }\href@noop {} {\bibfield  {journal} {\bibinfo  {journal} {Phys. Rev. Lett.}\
  }\textbf {\bibinfo {volume} {97}},\ p.\ \bibinfo {pages} {121801} (\bibinfo
  {year} {2006}{\natexlab{a}})}\BibitemShut {NoStop}%
\bibitem [{\citenamefont {Ablikim}\ \emph
  {et~al.}(2006{\natexlab{b}})\citenamefont {Ablikim} \emph {et~al.}}]{bib7}%
  \BibitemOpen
  \bibfield  {author} {\bibinfo {author} {\bibfnamefont {M.}~\bibnamefont
  {Ablikim}} \emph {et~al.} (\bibinfo {collaboration} {BES Collaboration}),\
  }\href@noop {} {\bibfield  {journal} {\bibinfo  {journal} {Phys. Lett. B}\
  }\textbf {\bibinfo {volume} {641}},\ p.\ \bibinfo {pages} {145} (\bibinfo
  {year} {2006}{\natexlab{b}})}\BibitemShut {NoStop}%
\bibitem [{\citenamefont {Besson}\ \emph {et~al.}(2010)\citenamefont {Besson}
  \emph {et~al.}}]{bib8}%
  \BibitemOpen
  \bibfield  {author} {\bibinfo {author} {\bibfnamefont {D.}~\bibnamefont
  {Besson}} \emph {et~al.} (\bibinfo {collaboration} {CLEO Collaboration}),\
  }\href@noop {} {\bibfield  {journal} {\bibinfo  {journal} {Phys. Rev. Lett.}\
  }\textbf {\bibinfo {volume} {104}},\ p.\ \bibinfo {pages} {159901(E)}
  (\bibinfo {year} {2010})}\BibitemShut {NoStop}%
\bibitem [{\citenamefont {Bai}\ \emph {et~al.}(2005)\citenamefont {Bai} \emph
  {et~al.}}]{bib9}%
  \BibitemOpen
  \bibfield  {author} {\bibinfo {author} {\bibfnamefont {J.~Z.}\ \bibnamefont
  {Bai}} \emph {et~al.} (\bibinfo {collaboration} {BES Collaboration}),\
  }\href@noop {} {\bibfield  {journal} {\bibinfo  {journal} {Phys. Lett. B}\
  }\textbf {\bibinfo {volume} {605}},\ p.~\bibinfo {pages} {63} (\bibinfo
  {year} {2005})}\BibitemShut {NoStop}%
\bibitem [{\citenamefont {Adam}\ \emph {et~al.}(2006)\citenamefont {Adam} \emph
  {et~al.}}]{bib10}%
  \BibitemOpen
  \bibfield  {author} {\bibinfo {author} {\bibfnamefont {N.~E.}\ \bibnamefont
  {Adam}} \emph {et~al.} (\bibinfo {collaboration} {CLEO Collaboration}),\
  }\href@noop {} {\bibfield  {journal} {\bibinfo  {journal} {Phys. Rev. Lett.}\
  }\textbf {\bibinfo {volume} {96}},\ p.\ \bibinfo {pages} {082004} (\bibinfo
  {year} {2006})}\BibitemShut {NoStop}%
\bibitem [{\citenamefont {Coan}\ \emph {et~al.}(2006)\citenamefont {Coan} \emph
  {et~al.}}]{bib11}%
  \BibitemOpen
  \bibfield  {author} {\bibinfo {author} {\bibfnamefont {T.~E.}\ \bibnamefont
  {Coan}} \emph {et~al.} (\bibinfo {collaboration} {CLEO Collaboration}),\
  }\href@noop {} {\bibfield  {journal} {\bibinfo  {journal} {Phys. Rev. Lett.}\
  }\textbf {\bibinfo {volume} {96}},\ p.\ \bibinfo {pages} {182002} (\bibinfo
  {year} {2006})}\BibitemShut {NoStop}%
\bibitem [{\citenamefont {Briere}\ \emph {et~al.}(2006)\citenamefont {Briere}
  \emph {et~al.}}]{bib12}%
  \BibitemOpen
  \bibfield  {author} {\bibinfo {author} {\bibfnamefont {R.~A.}\ \bibnamefont
  {Briere}} \emph {et~al.} (\bibinfo {collaboration} {CLEO Collaboration}),\
  }\href@noop {} {\bibfield  {journal} {\bibinfo  {journal} {Phys. Rev. D}\
  }\textbf {\bibinfo {volume} {74}},\ p.\ \bibinfo {pages} {031106} (\bibinfo
  {year} {2006})}\BibitemShut {NoStop}%
\bibitem [{\citenamefont {Adams}\ \emph {et~al.}(2006)\citenamefont {Adams}
  \emph {et~al.}}]{bib13}%
  \BibitemOpen
  \bibfield  {author} {\bibinfo {author} {\bibfnamefont {G.~S.}\ \bibnamefont
  {Adams}} \emph {et~al.} (\bibinfo {collaboration} {CLEO Collaboration}),\
  }\href@noop {} {\bibfield  {journal} {\bibinfo  {journal} {Phys. Rev. D}\
  }\textbf {\bibinfo {volume} {73}},\ p.\ \bibinfo {pages} {012002} (\bibinfo
  {year} {2006})}\BibitemShut {NoStop}%
\bibitem [{\citenamefont {Olive}\ \emph {et~al.}(2014)\citenamefont {Olive}
  \emph {et~al.}}]{pdg}%
  \BibitemOpen
  \bibfield  {author} {\bibinfo {author} {\bibfnamefont {K.~A.}\ \bibnamefont
  {Olive}} \emph {et~al.} (\bibinfo {collaboration} {Particle Data Group}),\
  }\href {\doibase 10.1088/1674-1137/38/9/090001} {\bibfield  {journal}
  {\bibinfo  {journal} {Chin. Phys. C}\ }\textbf {\bibinfo {volume} {38}},\ p.\
  \bibinfo {pages} {090001} (\bibinfo {year} {2014})}\BibitemShut {NoStop}%
\bibitem [{\citenamefont {Li}\ and\ \citenamefont {Zhao}(2011)}]{bib15}%
  \BibitemOpen
  \bibfield  {author} {\bibinfo {author} {\bibfnamefont {G.}~\bibnamefont
  {Li}}\ and\ \bibinfo {author} {\bibfnamefont {Q.}~\bibnamefont {Zhao}},\
  }\href@noop {} {\bibfield  {journal} {\bibinfo  {journal} {Phys. Rev. D}\
  }\textbf {\bibinfo {volume} {84}},\ p.\ \bibinfo {pages} {074005} (\bibinfo
  {year} {2011})}\BibitemShut {NoStop}%
\bibitem [{\citenamefont {Ablikim}\ \emph
  {et~al.}(2014{\natexlab{a}})\citenamefont {Ablikim} \emph
  {et~al.}}]{bibetac}%
  \BibitemOpen
  \bibfield  {author} {\bibinfo {author} {\bibfnamefont {M.}~\bibnamefont
  {Ablikim}} \emph {et~al.} (\bibinfo {collaboration} {BESIII Collaboration}),\
  }\href@noop {} {\bibfield  {journal} {\bibinfo  {journal} {Phys. Rev. D}\
  }\textbf {\bibinfo {volume} {89}},\ p.\ \bibinfo {pages} {112005} (\bibinfo
  {year} {2014}{\natexlab{a}})}\BibitemShut {NoStop}%
\bibitem [{\citenamefont {J.~J.~Dudek}\ and\ \citenamefont
  {Thomas}(2009)}]{bib16}%
  \BibitemOpen
  \bibfield  {author} {\bibinfo {author} {\bibfnamefont {R.~E.}\ \bibnamefont
  {J.~J.~Dudek}}\ and\ \bibinfo {author} {\bibfnamefont {C.~E.}\ \bibnamefont
  {Thomas}},\ }\href@noop {} {\bibfield  {journal} {\bibinfo  {journal} {Phys.
  Rev. D}\ }\textbf {\bibinfo {volume} {79}},\ p.\ \bibinfo {pages} {094504}
  (\bibinfo {year} {2009})}\BibitemShut {NoStop}%
\bibitem [{\citenamefont {Y.~B.~Ding}\ and\ \citenamefont
  {Chao}(1991)}]{bib17}%
  \BibitemOpen
  \bibfield  {author} {\bibinfo {author} {\bibfnamefont {D.~H.~Q.}\
  \bibnamefont {Y.~B.~Ding}}\ and\ \bibinfo {author} {\bibfnamefont {K.~T.}\
  \bibnamefont {Chao}},\ }\href@noop {} {\bibfield  {journal} {\bibinfo
  {journal} {Phys. Rev. D}\ }\textbf {\bibinfo {volume} {44}},\ p.\ \bibinfo
  {pages} {3562} (\bibinfo {year} {1991})}\BibitemShut {NoStop}%
\bibitem [{\citenamefont {Rosner}(2001{\natexlab{b}})}]{bib18}%
  \BibitemOpen
  \bibfield  {author} {\bibinfo {author} {\bibfnamefont {J.~L.}\ \bibnamefont
  {Rosner}},\ }\href@noop {} {\bibfield  {journal} {\bibinfo  {journal} {Phys.
  Rev. D}\ }\textbf {\bibinfo {volume} {64}},\ p.\ \bibinfo {pages} {094002}
  (\bibinfo {year} {2001}{\natexlab{b}})}\BibitemShut {NoStop}%
\bibitem [{\citenamefont {E.~J.~Eichten}\ and\ \citenamefont
  {Quigg}(2004)}]{bib19}%
  \BibitemOpen
  \bibfield  {author} {\bibinfo {author} {\bibfnamefont {K.~L.}\ \bibnamefont
  {E.~J.~Eichten}}\ and\ \bibinfo {author} {\bibfnamefont {C.}~\bibnamefont
  {Quigg}},\ }\href@noop {} {\bibfield  {journal} {\bibinfo  {journal} {Phys.
  Rev. D}\ }\textbf {\bibinfo {volume} {69}},\ p.\ \bibinfo {pages} {094019}
  (\bibinfo {year} {2004})}\BibitemShut {NoStop}%
\bibitem [{\citenamefont {Ablikim}\ \emph
  {et~al.}(2015{\natexlab{a}})\citenamefont {Ablikim} \emph
  {et~al.}}]{bibchicJ}%
  \BibitemOpen
  \bibfield  {author} {\bibinfo {author} {\bibfnamefont {M.}~\bibnamefont
  {Ablikim}} \emph {et~al.} (\bibinfo {collaboration} {BESIII Collaboration}),\
  }\href@noop {} {\bibfield  {journal} {\bibinfo  {journal} {Phys. Rev. D}\
  }\textbf {\bibinfo {volume} {91}},\ p.\ \bibinfo {pages} {092009} (\bibinfo
  {year} {2015}{\natexlab{a}})}\BibitemShut {NoStop}%
\bibitem [{\citenamefont {F.~K.~Guo}\ and\ \citenamefont {Zhao}(2010)}]{bib25}%
  \BibitemOpen
  \bibfield  {author} {\bibinfo {author} {\bibfnamefont {G.~L. U. G.~M.}\
  \bibnamefont {F.~K.~Guo}, \bibfnamefont {C.~Hanhart}}\ and\ \bibinfo {author}
  {\bibfnamefont {Q.}~\bibnamefont {Zhao}},\ }\href@noop {} {\bibfield
  {journal} {\bibinfo  {journal} {Phys. Rev. D}\ }\textbf {\bibinfo {volume}
  {82}},\ p.\ \bibinfo {pages} {034025} (\bibinfo {year} {2010})}\BibitemShut
  {NoStop}%
\bibitem [{\citenamefont {Ablikim}\ \emph
  {et~al.}(2015{\natexlab{b}})\citenamefont {Ablikim} \emph
  {et~al.}}]{bibpi0etac}%
  \BibitemOpen
  \bibfield  {author} {\bibinfo {author} {\bibfnamefont {M.}~\bibnamefont
  {Ablikim}} \emph {et~al.} (\bibinfo {collaboration} {BESIII Collaboration}),\
  }\href@noop {} {\bibfield  {journal} {\bibinfo  {journal} {Phys. Rev. D}\
  }\textbf {\bibinfo {volume} {91}},\ p.\ \bibinfo {pages} {112018} (\bibinfo
  {year} {2015}{\natexlab{b}})}\BibitemShut {NoStop}%
\bibitem [{\citenamefont {Ablikim}\ \emph
  {et~al.}(2014{\natexlab{b}})\citenamefont {Ablikim} \emph
  {et~al.}}]{bibgggphi}%
  \BibitemOpen
  \bibfield  {author} {\bibinfo {author} {\bibfnamefont {M.}~\bibnamefont
  {Ablikim}} \emph {et~al.} (\bibinfo {collaboration} {BESIII Collaboration}),\
  }\href {\doibase 10.1103/PhysRevD.90.092002} {\bibfield  {journal} {\bibinfo
  {journal} {Phys. Rev. D}\ }\textbf {\bibinfo {volume} {90}},\ p.\ \bibinfo
  {pages} {092002} (\bibinfo {year} {2014}{\natexlab{b}})}\BibitemShut
  {NoStop}%
\bibitem [{\citenamefont {Adams}\ \emph {et~al.}(2008)\citenamefont {Adams}
  \emph {et~al.}}]{bib26}%
  \BibitemOpen
  \bibfield  {author} {\bibinfo {author} {\bibfnamefont {G.~S.}\ \bibnamefont
  {Adams}} \emph {et~al.} (\bibinfo {collaboration} {CLEO Collaboration}),\
  }\href@noop {} {\bibfield  {journal} {\bibinfo  {journal} {Phys. Rev. Lett.}\
  }\textbf {\bibinfo {volume} {101}},\ p.\ \bibinfo {pages} {101801} (\bibinfo
  {year} {2008})}\BibitemShut {NoStop}%
\bibitem [{\citenamefont {Haber}\ and\ \citenamefont {Perrier}(1985)}]{bib27}%
  \BibitemOpen
  \bibfield  {author} {\bibinfo {author} {\bibfnamefont {H.~E.}\ \bibnamefont
  {Haber}}\ and\ \bibinfo {author} {\bibfnamefont {J.}~\bibnamefont
  {Perrier}},\ }\href@noop {} {\bibfield  {journal} {\bibinfo  {journal} {Phys.
  Rev. D}\ }\textbf {\bibinfo {volume} {32}},\ p.\ \bibinfo {pages} {2961}
  (\bibinfo {year} {1985})}\BibitemShut {NoStop}%
\bibitem [{\citenamefont {A.~Seiden}\ and\ \citenamefont
  {Haber}(1988)}]{bib28}%
  \BibitemOpen
  \bibfield  {author} {\bibinfo {author} {\bibfnamefont {H.~F. W.~S.}\
  \bibnamefont {A.~Seiden}}\ and\ \bibinfo {author} {\bibfnamefont {H.~E.}\
  \bibnamefont {Haber}},\ }\href@noop {} {\bibfield  {journal} {\bibinfo
  {journal} {Phys. Rev. D}\ }\textbf {\bibinfo {volume} {38}},\ p.\ \bibinfo
  {pages} {824} (\bibinfo {year} {1988})}\BibitemShut {NoStop}%
\bibitem [{\citenamefont {Escribano}\ \emph {et~al.}(2010)\citenamefont
  {Escribano} \emph {et~al.}}]{bib29}%
  \BibitemOpen
  \bibfield  {author} {\bibinfo {author} {\bibfnamefont {R.}~\bibnamefont
  {Escribano}} \emph {et~al.} (\bibinfo {collaboration} {CLEO Collaboration}),\
  }\href@noop {} {\bibfield  {journal} {\bibinfo  {journal} {Eur. Phys. J. C}\
  }\textbf {\bibinfo {volume} {65}},\ p.\ \bibinfo {pages} {467} (\bibinfo
  {year} {2010})}\BibitemShut {NoStop}%
\bibitem [{\citenamefont {Ablikim}\ \emph
  {et~al.}(2015{\natexlab{c}})\citenamefont {Ablikim} \emph {et~al.}}]{bib30}%
  \BibitemOpen
  \bibfield  {author} {\bibinfo {author} {\bibfnamefont {M.}~\bibnamefont
  {Ablikim}} \emph {et~al.} (\bibinfo {collaboration} {BESIII Collaboration}),\
  }\href {\doibase 10.1103/PhysRevD.91.112001} {\bibfield  {journal} {\bibinfo
  {journal} {Phys. Rev. D}\ }\textbf {\bibinfo {volume} {91}},\ p.\ \bibinfo
  {pages} {112001} (\bibinfo {year} {2015}{\natexlab{c}})}\BibitemShut
  {NoStop}%
\bibitem [{\citenamefont {Ablikim}\ \emph {et~al.}(2012)\citenamefont {Ablikim}
  \emph {et~al.}}]{bib1405}%
  \BibitemOpen
  \bibfield  {author} {\bibinfo {author} {\bibfnamefont {M.}~\bibnamefont
  {Ablikim}} \emph {et~al.} (\bibinfo {collaboration} {BESIII Collaboration}),\
  }\href {\doibase 10.1103/PhysRevLett.108.182001} {\bibfield  {journal}
  {\bibinfo  {journal} {Phys. Rev. Lett.}\ }\textbf {\bibinfo {volume} {108}},\
  p.\ \bibinfo {pages} {182001} (\bibinfo {year} {2012})}\BibitemShut {NoStop}%
\bibitem [{\citenamefont {Wu}\ \emph {et~al.}(2012)\citenamefont {Wu},
  \citenamefont {Liu}, \citenamefont {Zhao},\ and\ \citenamefont {Zou}}]{tsm1}%
  \BibitemOpen
  \bibfield  {author} {\bibinfo {author} {\bibfnamefont {J.~J.}\ \bibnamefont
  {Wu}}, \bibinfo {author} {\bibfnamefont {X.~H.}\ \bibnamefont {Liu}},
  \bibinfo {author} {\bibfnamefont {Q.}~\bibnamefont {Zhao}}, \ and\ \bibinfo
  {author} {\bibfnamefont {B.~S.}\ \bibnamefont {Zou}},\ }\href {\doibase
  10.1103/PhysRevLett.108.081803} {\bibfield  {journal} {\bibinfo  {journal}
  {Phys. Rev. Lett.}\ }\textbf {\bibinfo {volume} {108}},\ p.\ \bibinfo {pages}
  {081803} (\bibinfo {year} {2012})}\BibitemShut {NoStop}%
\bibitem [{\citenamefont {Aceti}\ \emph {et~al.}(2012)\citenamefont {Aceti},
  \citenamefont {Liang}, \citenamefont {Oset}, \citenamefont {Wu},\ and\
  \citenamefont {Zou}}]{tsm2}%
  \BibitemOpen
  \bibfield  {author} {\bibinfo {author} {\bibfnamefont {F.}~\bibnamefont
  {Aceti}}, \bibinfo {author} {\bibfnamefont {W.~H.}\ \bibnamefont {Liang}},
  \bibinfo {author} {\bibfnamefont {E.}~\bibnamefont {Oset}}, \bibinfo {author}
  {\bibfnamefont {J.~J.}\ \bibnamefont {Wu}}, \ and\ \bibinfo {author}
  {\bibfnamefont {B.~S.}\ \bibnamefont {Zou}},\ }\href {\doibase
  10.1103/PhysRevD.86.114007} {\bibfield  {journal} {\bibinfo  {journal} {Phys.
  Rev. D}\ }\textbf {\bibinfo {volume} {86}},\ p.\ \bibinfo {pages} {114007}
  (\bibinfo {year} {2012})}\BibitemShut {NoStop}%
\bibitem [{\citenamefont {Ablikim}\ \emph
  {et~al.}(2015{\natexlab{d}})\citenamefont {Ablikim} \emph
  {et~al.}}]{bib14052}%
  \BibitemOpen
  \bibfield  {author} {\bibinfo {author} {\bibfnamefont {M.}~\bibnamefont
  {Ablikim}} \emph {et~al.} (\bibinfo {collaboration} {BESIII Collaboration}),\
  }\href {\doibase 10.1103/PhysRevD.92.012007} {\bibfield  {journal} {\bibinfo
  {journal} {Phys. Rev. D}\ }\textbf {\bibinfo {volume} {92}},\ p.\ \bibinfo
  {pages} {012007} (\bibinfo {year} {2015}{\natexlab{d}})}\BibitemShut
  {NoStop}%
\bibitem [{\citenamefont {Aceti}, \citenamefont {Dias},\ and\ \citenamefont
  {Oset}(2015)}]{bibtsm3}%
  \BibitemOpen
  \bibfield  {author} {\bibinfo {author} {\bibfnamefont {F.}~\bibnamefont
  {Aceti}}, \bibinfo {author} {\bibfnamefont {J.~M.}\ \bibnamefont {Dias}}, \
  and\ \bibinfo {author} {\bibfnamefont {E.}~\bibnamefont {Oset}},\ }\href
  {\doibase 10.1140/epja/i2015-15048-5} {\bibfield  {journal} {\bibinfo
  {journal} {Eur. Phys. J. A}\ }\textbf {\bibinfo {volume} {51}},\ p.~\bibinfo
  {pages} {48} (\bibinfo {year} {2015})}\BibitemShut {NoStop}%
\bibitem [{\citenamefont {Bonvicini}\ \emph {et~al.}(2008)\citenamefont
  {Bonvicini} \emph {et~al.}}]{etapk1}%
  \BibitemOpen
  \bibfield  {author} {\bibinfo {author} {\bibfnamefont {G.}~\bibnamefont
  {Bonvicini}} \emph {et~al.} (\bibinfo {collaboration} {CLEO Collaboration}),\
  }\href {\doibase 10.1103/PhysRevD.78.052001} {\bibfield  {journal} {\bibinfo
  {journal} {Phys. Rev. D}\ }\textbf {\bibinfo {volume} {78}},\ p.\ \bibinfo
  {pages} {052001} (\bibinfo {year} {2008})}\BibitemShut {NoStop}%
\bibitem [{\citenamefont {Bugg}(2006)}]{etapk2}%
  \BibitemOpen
  \bibfield  {author} {\bibinfo {author} {\bibfnamefont {D.~V.}\ \bibnamefont
  {Bugg}},\ }\href {\doibase 10.1016/j.physletb.2005.11.019} {\bibfield
  {journal} {\bibinfo  {journal} {Phys. Lett. B}\ }\textbf {\bibinfo {volume}
  {632}},\ p.\ \bibinfo {pages} {471} (\bibinfo {year} {2006})}\BibitemShut
  {NoStop}%
\bibitem [{\citenamefont {Ablikim}\ \emph
  {et~al.}(2014{\natexlab{c}})\citenamefont {Ablikim} \emph {et~al.}}]{etapkk}%
  \BibitemOpen
  \bibfield  {author} {\bibinfo {author} {\bibfnamefont {M.}~\bibnamefont
  {Ablikim}} \emph {et~al.} (\bibinfo {collaboration} {BESIII Collaboration}),\
  }\href {\doibase 10.1103/PhysRevD.89.074030} {\bibfield  {journal} {\bibinfo
  {journal} {Phys. Rev. D}\ }\textbf {\bibinfo {volume} {89}},\ p.\ \bibinfo
  {pages} {074030} (\bibinfo {year} {2014}{\natexlab{c}})}\BibitemShut
  {NoStop}%
\bibitem [{\citenamefont {D.~M.~Li}\ and\ \citenamefont {Yu}(2005)}]{13801}%
  \BibitemOpen
  \bibfield  {author} {\bibinfo {author} {\bibfnamefont {B.~M.}\ \bibnamefont
  {D.~M.~Li}}\ and\ \bibinfo {author} {\bibfnamefont {H.}~\bibnamefont {Yu}},\
  }\href@noop {} {\bibfield  {journal} {\bibinfo  {journal} {Eur. Phys. J.
  direct A}\ }\textbf {\bibinfo {volume} {26}},\ p.\ \bibinfo {pages} {141}
  (\bibinfo {year} {2005})}\BibitemShut {NoStop}%
\bibitem [{\citenamefont {Godfrey}\ and\ \citenamefont {Isgur}(1985)}]{13802}%
  \BibitemOpen
  \bibfield  {author} {\bibinfo {author} {\bibfnamefont {S.}~\bibnamefont
  {Godfrey}}\ and\ \bibinfo {author} {\bibfnamefont {N.}~\bibnamefont
  {Isgur}},\ }\href {\doibase 10.1103/PhysRevD.32.189} {\bibfield  {journal}
  {\bibinfo  {journal} {Phys. Rev. D}\ }\textbf {\bibinfo {volume} {32}},\ p.\
  \bibinfo {pages} {189} (\bibinfo {year} {1985})}\BibitemShut {NoStop}%
\bibitem [{\citenamefont {Ablikim}\ \emph
  {et~al.}(2015{\natexlab{e}})\citenamefont {Ablikim} \emph {et~al.}}]{k892}%
  \BibitemOpen
  \bibfield  {author} {\bibinfo {author} {\bibfnamefont {M.}~\bibnamefont
  {Ablikim}} \emph {et~al.} (\bibinfo {collaboration} {BESIII Collaboration}),\
  }\href {\doibase 10.1103/PhysRevD.91.112008} {\bibfield  {journal} {\bibinfo
  {journal} {Phys. Rev. D}\ }\textbf {\bibinfo {volume} {91}},\ p.\ \bibinfo
  {pages} {112008} (\bibinfo {year} {2015}{\natexlab{e}})}\BibitemShut
  {NoStop}%
\end{thebibliography}%

\end{document}